# A Co-simulation Framework for Quadrotor Control System Design using ROS 2 and MATLAB/Simulink

Moscow Aviation Institution

by Hangyu Teng

**Abstract**—— Co-simulation is a critical approach for the design and analysis of complex cyber-physical systems. It will enhance development efficiency and reduce costs. This paper presents a co-simulation framework integrating ROS 2 and MATLAB/Simulink for quadrotor unmanned aerial vehicle (UAV) control system design and verification. First, a six-degree-of-freedom nonlinear dynamic model of the quadrotor is derived accurately that based on Newton-Euler equations. Second, within the proposed framework, a hierarchical control architecture was designed and implemented: LQR controller for attitude control to achieve optimal regulation performance, and PID controller for position control to ensure robustness and practical applicability. Third, elaborated the architecture of the framework, including the implementation details of the cross-platform data exchange mechanism. Simulation results demonstrate the effectiveness of the framework, highlighting its capability to provide an efficient and standardized solution for rapid prototyping and Software-in-the-Loop (SIL) validation of UAV control algorithms.

## I. INTRODUCTION

Recent years, the quadrotor UAV has extensive application in both of civil and military fields because its simple structure, flexible maneuverability and low cost. From aerial photography, logistics and delivery, agricultural plant protection, to search and rescue, environmental monitoring, and infrastructure inspection, quadrotors are profoundly transforming the traditional industries by reshape their working model. However, as a highly coupled, underactuated and nonlinear system, design and verification of quadrotor control system is still a core problem faced by academia and industry fields. First, the system possesses six degrees of freedom but only four controllable outputs, belongs underactuated system, position control needs to be achieved indirectly through attitude control; secondly, system dynamic presents strongly nonlinearity, especially when maneuvers at a large angle, linearized model want to describe system behavior accurately is very difficult. Thirdly, the system has multi-variable coupled effect, pitching, rolling and heading motions will influence each other, thereby increasing complexity of controller design; Finally, practical system must consider modeling errors, sensors noise, actuator delays and external disturbances. These factors contribute to a prominent gap between theoretical design and practical application.

Traditional control system follows a linear development process that usually form design to simulation to prototype to verify. However, this approach has obvious drawbacks that is it can't capture the complex behavior of practical system, because it usually depends on simplified mathematic model during design phase; simulation verification is separated from the actual test environment, cause simulation succeed but actual test fail; moreover, manufacturing hardware prototypes have a long period, high cost, and security risk; and the feedback of iteration is slow, as soon as finding problems we need to embark on a redesign, it will prolong development cycle. Therefore, there is an urgent need for a new development paradigm that can shorten the design cycle, reduce development costs and enhance verification reliability.

Co-simulation as an emerging system engineering method, supply an effective way to solve the above question. The core idea is decomposing the complex system to several subsystem, then put the subsystem into different simulation environment which should be suited them to model and solve. Implement data exchange and time synchronization by standardized interfaces, thereby achieved co-simulation of whole system. This paper selects ROS 2 Jazzy (Robot Operating System 2）and MATLAB/Simulink as tool for co-simulation, the reason lies in them two formed a perfect complement:

(1) **Field professionalism**: Simulink is the industry standard for design of control algorithm, with their graphical modeling and a rich mathematical toolbox; meanwhile ROS 2 provides a distributed communication architecture and rich robot function packages, making it a preferred platform for system collection and soft framework development.

(2) **Real-Time Performance**: ROS 2 is based on the DDS (Data Distribution Service) middleware and can meet the hard real-time requirements of the control system through QoS policy configuration.

(3) **Cross-platform and Multi-language collaboration**: ROS 2 supports Linux, Windows and embedded systems, with compatible for multiple program languages including C++ and Python, it's convenient to parallel develop using diverse tools; MATLAB can provide a robust support for modeling complex nonlinear system by its advanced symbolic computation and numerical analysis capability. Additionally, Simulink Coder can transform to the optimized C/C++ code automatically, supports seamless transition from simulation to embedded deployment.

The descriptions of subsequent sections are as follows: section 2 and section 3 will respectively introduce dynamic model of system and designing method of controller. Section 4 will elaborate architecture of co-simulation and its implementation. Section 5 will show the crucial simulation results and performance analysis. Finally, section 6 will conclude the paper and outlines potential future directions.

## II. QUADROTOR SYSTEM MODELLING AND CONTROL

(1) Principles of Quadcopter Operation

A quadrotor consists of four symmetrically distributed rotors, with typical configurations being the 'X' form and the '+' form. The total thrust of the quadrotor is the sum of all four rotor thrusts:

$$F_{total} = F_1 + F_2 + F_3 + F_4 = k_t(\omega_1^2 + \omega_2^2 + \omega_3^2 + \omega_4^2) \quad (1)$$

where $F_i$ is the thrust force generated by the *i*-th rotor (N), $\omega_i$ is the angular velocity of the *i*-th motor (*rad/s*), and $k_t$ is the thrust coefficient (*N·s²/rad²*). In hover or steady flight, the total reactive torque is balanced:

$$\tau_{total} = k_m(\omega_1^2 + \omega_3^2) - k_m(\omega_2^2 + \omega_4^2) = 0 \quad (2)$$

Where $k_m$ is the torque coefficient (*N·m·s²/rad²*). The torque direction follows the right-hand rule.

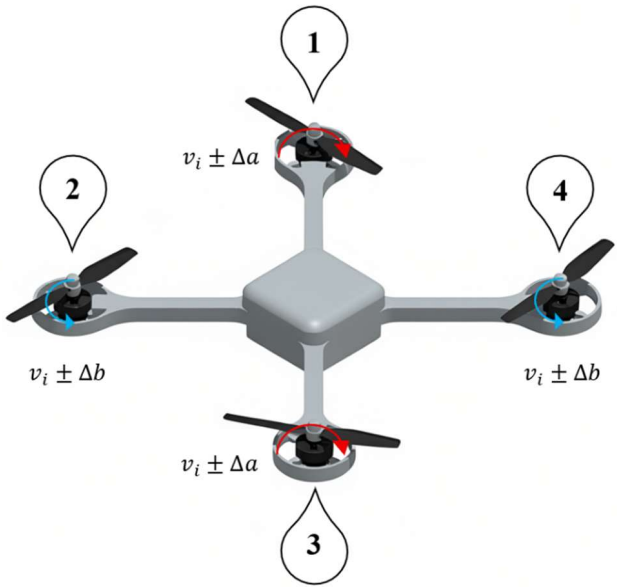

**Figure 1 Principles of Quadcopter Operation**

A quadrotor achieves six-degree-of-freedom motion control by differentially adjusting the rotational speeds of its four motors. As illustrated in Fig. 1, the operating principles demonstrate how motors control altitude and attitude through varying speeds. Before proceeding, we must define the coordinate frames.

- **Earth-fixed Inertial Frame** $\mathcal{F}_E = \{O_E; x_E, y_E, z_E\}$: A fixed reference frame attached to the ground, used to describe the absolute position and attitude of the vehicle. We adopt the standard ENU (East-North-Up) convention, where $x_E$ points east, $y_E$ points north, $z_E$ and points vertically upward.
- **Body-fixed Frame** $\mathcal{F}_B = \{O_B; x_B, y_B, z_B\}$: A moving reference frame attached to the quadrotor's center of mass, where $x_B$ points toward the nose (rotor 1 direction), $y_B$ points to the left side (rotor 4 direction), and $z_B$ points vertically upward perpendicular to the body plane. The vehicle's linear velocity and angular velocity in the body frame are represented by vectors $v_b = [u, v, w]^T$ and $\omega_b = [p, q, r]^T$, respectively.

Let $v_i$ denote the nominal hovering speed of each motor. Define $\Delta a$ and $\Delta b$ as the differential speed adjustments applied to opposing motor pairs. The motion modes are determined as follows:

- **Altitude control**: when $\Delta a = \Delta b > 0 \Rightarrow F_{total} > mg$ vertical ascent; conversely, causes descent.
- **Roll motion**: when $\Delta a = 0, \Delta b \neq 0$ achieved roll.
- **Pitch motion**: when $\Delta a \neq 0, \Delta b = 0$ achieved pitch.
- **Yaw motion**: when $\Delta a \neq \Delta b \neq 0$ achieved yaw.

The attitude of quadrotor is defined by a set of Euler angles: $\Phi = [\phi, \theta, \psi]^T$ represent roll angle, pitch angle and yaw angle respectively. The transformation of attitude from the ground frame to the body frame can be achieved using a rotation matrix $C_{fru/enu}$, the sequence of rotations is Z-Y-X.

$$C_{fru/enu} = \begin{bmatrix} c\theta c\psi & c\theta s\psi & -s\theta \\ s\phi s\theta c\psi - c\phi s\psi & s\phi s\theta s\psi + c\phi c\psi & s\phi c\theta \\ c\phi s\theta c\psi + s\phi s\psi & c\phi s\theta s\psi - s\phi c\psi & c\phi c\theta \end{bmatrix} \quad (3)$$

Here, $c(\cdot)$ and $s(\cdot)$ denote $cos(\cdot)$ and $sin(\cdot)$, respectively.

(2) Thrust and Torque Model

The motion of a quadrotor is generated by four independently controlled brushless DC motor-driven propellers.

$$\begin{bmatrix} U_1 \\ U_2 \\ U_3 \\ U_4 \end{bmatrix} = \begin{bmatrix} k_f & k_f & k_f & k_f \\ 0 & -Lk_f & 0 & Lk_f \\ -Lk_f & 0 & Lk_f & 0 \\ k_M & -k_M & k_M & -k_M \end{bmatrix} \begin{bmatrix} \omega_1^2 \\ \omega_2^2 \\ \omega_3^2 \\ \omega_4^2 \end{bmatrix} \quad (4)$$

To facilitate controller design, we combine the individual effects of the four motors into a total thrust vector $U_1$ and three control moments $U_2, U_3, U_4$ acting on the body axes. This mapping relationship is defined by the mixing matrix (4). Here, L is the distance from rotors to center of mass

(3) Equations of Motion

Using the Newton-Euler method, we can derive the dynamic equations that describe rigid body motion.

i) Translational Dynamics:

$$\dot{v}_{cm/e}^{fru} = \frac{F_b}{m} + C_{fru/enu} G_e \tag{5}$$

$$\dot{p}_{cm/e}^{enu} = C_{enu/fru} v_{cm/e}^{fru} \tag{6}$$

Here, m denotes the total mass of the quadrotor. $F_b = [0,0,U_1]^T$; $G_e = [0,0,-g]^T$; the gyroscopic coupling terms are negligible compared to the dominant thrust and gravitational forces.

$$m \begin{bmatrix} \dot{u} \\ \dot{v} \\ \dot{w} \end{bmatrix} = \begin{bmatrix} 0 \\ 0 \\ U_1 \end{bmatrix} + \begin{bmatrix} mg\sin\theta \\ -mg\cos\theta\sin\phi \\ -mg\cos\theta\cos\phi \end{bmatrix}$$

By substituting Equations (4)(6) and expanding it, we obtain the translational equations of motion in component form equation (7):

$$\begin{cases} \ddot{x}_e = \frac{U_1}{m}(\cos\psi\sin\theta\cos\phi + \sin\psi\sin\phi) \\ \ddot{y}_e = \frac{U_1}{m}(\sin\psi\sin\theta\cos\phi - \cos\psi\sin\phi) \\ \ddot{z}_e = \frac{U_1}{m}(\cos\theta\cos\phi) - g \end{cases} \tag{7}$$

ii) Rotational Dynamics:

We describe the rotation of quadrotor within Body-fixed reference frame:

$$\dot{\omega}_{b/e}^{bf} = (I^{bf})^{-1}[\bar{U} - \omega_{b/e}^{bf} \times I^{bf}\omega_{b/e}^{bf}] \tag{8}$$

Here, $\bar{U} = [U_2, U_3, U_4]^T$, the inertia matrix is equation (9), and because of symmetry this has a relatively simple form:

$$I^{bf} = \begin{bmatrix} I_{xx} & 0 & 0 \\ 0 & I_{yy} & 0 \\ 0 & 0 & I_{zz} \end{bmatrix} \tag{9}$$

Expanding equation (8), we derive the expression for angular acceleration.

$$\begin{cases} \dot{p} = \frac{1}{I_{xx}}(U_2 + (I_{yy} - I_{zz})qr) \\ \dot{q} = \frac{1}{I_{yy}}(U_3 + (I_{zz} - I_{xx})pr) \\ \dot{r} = \frac{1}{I_{zz}}(U_4 + (I_{xx} - I_{yy})pq) \end{cases} \tag{10}$$

The quadratic terms of angular velocity are higher-order small quantities, and the resulting gyroscopic coupling torque is negligible compared to the control torque. Therefore, Equation (10) can be reasonably simplified into the following linear decoupled form:

$$\begin{cases} \dot{p} = \frac{1}{I_{xx}} U_2 \\ \dot{q} = \frac{1}{I_{yy}} U_3 \\ \dot{r} = \frac{1}{I_{zz}} U_4 \end{cases} \tag{11}$$

Equations (7) and (11) collectively form the complete nonlinear dynamic model of the quadrotor. This model precisely describes how control inputs influence the motion of quadrotor, serving as the mathematical foundation for subsequent controller design and simulation.

(4) Controller Design

This section will design a hybrid control strategy for the quadrotor to achieve stable hovering and attitude tracking. For the faster inner attitude loop, we design an optimal Linear-Quadratic Regulator (LQR) to ensure dynamic regulation performance. Concurrently, for the outer position loop, a classic PID controller is designed to ensure robustness against external disturbances.

A) PD Controller Design for Altitude Channel

The altitude control objective is to guide the aircraft to reach and maintain a desired altitude $z_d$. We adopt a hybrid strategy that combines feedforward compensation with PD feedback control. To counteract gravity during hover($\theta = 0, \phi = 0$), require a baseline thrust $U_f = mg$. $U_f$ serves as the feedforward term, providing the primary lift input required by the system. Subsequently, we introduce a PD feedback controller to dynamically adjust thrust and eliminate altitude errors $e_z = z_d - z$. The PD control law is defined as:

$$U_{fb} = K_{p,z}e_z + K_{d,z}\dot{e}_z = K_{p,z}(z_{des} - z) - K_{d,z}\dot{z} \tag{12}$$

Where $K_p$ is proportional gain, $K_d$ is derivative gain. $\dot{z}_d = 0$. Finally, toral thrust $U_1$ is feedforward term plus feedback term as follows:

$$U_1 = \frac{U_f + U_{fb}}{\cos\theta\cos\phi} = \frac{mg + K_{p,z}(z_{des} - z) - K_{d,z}\dot{z}}{\cos\theta\cos\phi} \tag{13}$$

B) LQR Controller Design for Attitude Channels

Since the roll, pitch, and yaw dynamics of a quadrotor are approximately decoupled near the hover point, we can design independent controllers for each attitude channel. Here, we take the roll channel ($\phi$) as an illustrative example, while noting that the design process for pitch and yaw channels follows an identical symmetric approach.

First, transform the simplified dynamic model of the roll channel into state-space form. Define the state vector as $x_\phi =$

$[\phi, \dot{\phi}]^T$, tracking error is defined as $e_\phi = \phi_d - \phi$; input control is rolling moment $U_2$, state-space expression is as follows:

$$\begin{bmatrix} \dot{e}_\phi \\ \ddot{\phi} \end{bmatrix} = \begin{bmatrix} 0 & -1 \\ 0 & 0 \end{bmatrix} \begin{bmatrix} e_\phi \\ \dot{\phi} \end{bmatrix} + \begin{bmatrix} 0 \\ 1/I_{xx} \end{bmatrix} U_2 \quad (14)$$

Control law as follows:

$$U_2 = -\begin{bmatrix} k_1 & k_2 \end{bmatrix} \begin{bmatrix} e \\ \dot{\phi} \end{bmatrix} \quad (15)$$

We adopt the Linear Quadratic Regulator (LQR), an optimal control method, by solving the Algebraic Ricotti Equation (ARE) to obtain the optimal feedback gain matrix $[k_1, k_2]$. The cost function J is defined as:

$$J = \int_0^\infty \begin{bmatrix} e_\phi & \dot{\phi} \end{bmatrix} \cdot Q \cdot \begin{bmatrix} e_\phi \\ \dot{\phi} \end{bmatrix} + (U_2)^T R(U_2) dt \quad (16)$$

The controllers for pitch moment $U_3$ and yaw moment $U_4$ are independently designed using the same method, thereby forming a complete attitude control system. This LQR-based error feedback controller differs from conventional PID controllers in that its feedback gains are derived through a systematic optimization process, which theoretically yields superior dynamic performance.

III. CO-SIMULATION FRAMEWORK DESIGN

This section will elaborate a co-simulation architecture that assembles MATLAB/Simulink and ROS 2. This framework integrates Simulink's dynamic system modeling and simulation capabilities with ROS 2's modular and distributed robotic system, thereby providing an efficient and standard Software-in-the-Loop, SIL, verification environment for development of quadrotor control algorithms.

On the Simulink side, we have implemented a comprehensive closed-loop system that assemble both the dynamic model and an interface module for ROS 2 network communication, as illustrated in Figure 2.

- **ROS 2 Interface**: This serves as the communication bridge between the model and the external controller. A ROS 2 Subscriber module is responsible for subscribing to the */drone/motor_commands* topic to receive four-channel motor speed commands from the C++ controller. The received messages are parsed by an initialization module before being fed into the motor model.
- **Motor and Mixing Model**: This module transforms the input motor speed commands into forces acting on the airframe. It first computes individual thrusts proportional to the square of rotational speeds, then converts them via a mixing matrix (as specified in Equation (4)) into total thrust $U_1$ and triaxial control moments $U_2, U_3, U_4$.
- **6-DOF Dynamics Model**: Serving as the physical core of the simulation, this model comprises two submodules: Attitude Dynamics and Position Dynamics. The Attitude Dynamics submodule numerically integrates angular acceleration and angular velocity based on input control moments and inertial parameters to derive the fuselage's attitude angles $\Phi$ and angular rates $\omega_b$. The Position Dynamics submodule computes linear accelerations in the earth coordinate frame using total force, current attitude, and gravitational effects, then obtains linear velocity $v_b$ and position $x_E, y_E, z_E$ through analogous integration.
- **State Output Interface**: To send simulation states to external ROS 2 nodes, the model packages computed state variables (e.g., attitude angles and angular rates) into standard ROS 2 message formats. A critical procedure involves converting Euler angles to quaternions to circumvent gimbal lock issues. Ultimately, the model publishes simulated IMU data in compliant formats to designated topics via a ROS 2 Publisher module, thereby completing a simulation loop.

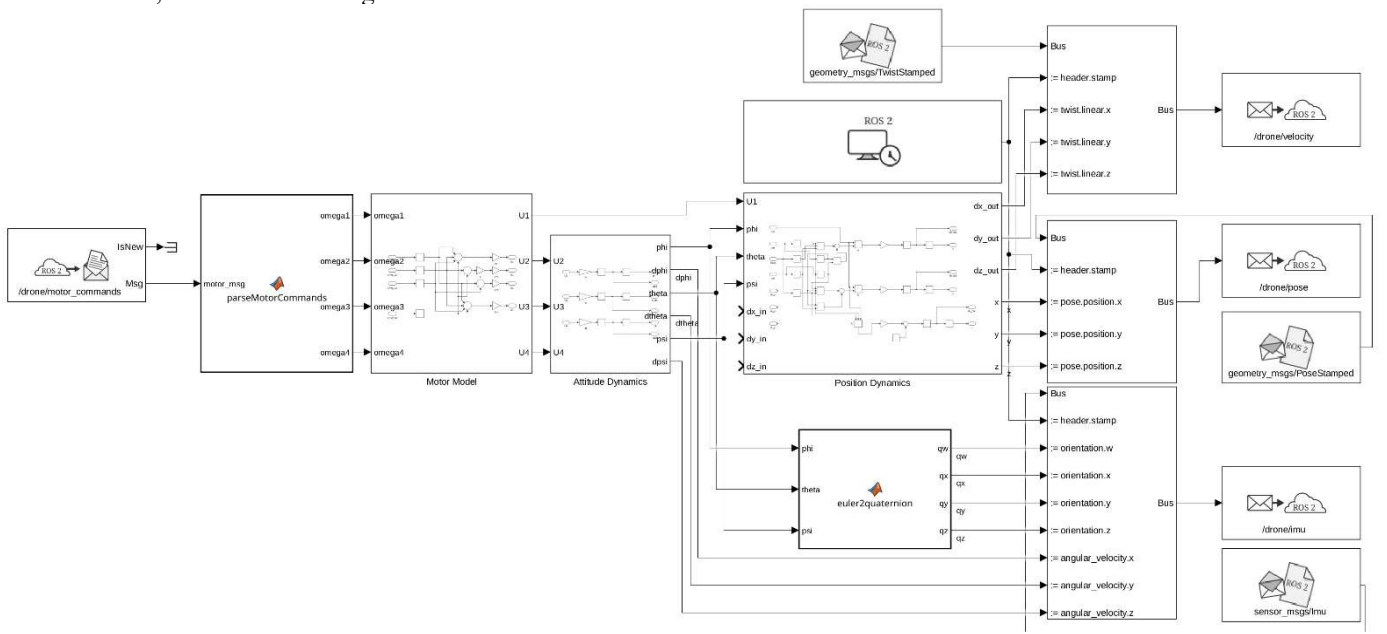

Figure 2 Block diagram of the Simulink model

Topics：
1. */drone/Velocity*: Publishes ROS 2 messages of type *geometry_msgs/TwistStamped*, containing the linear velocity data of the quadrotor。
2. */drone/Pose*: Publishes ROS 2 messages of type *geometry_msgs/PoseStamped*, which encapsulate the positional coordinates and orientation parameters of the quadrotor.
3. */drone/Imu*: Publishes ROS 2 messages of type *sensor_msgs/Imu*, emulating the output of an inertial measurement unit (IMU) sensor, which includes the UAV's orientation and angular velocity measurements.

On the ROS 2 side, this study presents the development of an *attitude_controller* package for ROS 2 on the Ubuntu platform. The core component is a C++ class named *AttitudeControllerNode*, which inherits from *rclcpp::Node* and encapsulates all control logic and communication interfaces. The node architecture adheres to ROS 2 best practices, utilizing the Parameter Server for managing controller gains and setpoints, while employing a Timer mechanism to ensure real-time performance and deterministic behavior of the control loop.

The core logic of this node can be summarized into the following key steps:

(1) Initialization: The node's constructor performs all necessary initialization tasks, including:

**Algorithm 1:** Quadrotor Controller Node Main Loop
**Data:**
Current state: $\mathbf{x} \leftarrow \{\phi, \theta, \psi, p, q, r, x, \dot{x}, y, \dot{y}, z, \dot{z}\}$
Desired state: $\mathbf{x}_{des} \leftarrow \{\phi_{des}, \theta_{des}, \psi_{des}, z_{des}\}$
**Result:**
Motor commands: $\mathbf{\Omega} = [\omega_1, \omega_2, \omega_3, \omega_4]^T$

1. **begin**
2.   **procedure** controlLoop()
3.     **if** sensor data is stale or missing **then**
4.       Publish safety motor commands;
5.       **return**;
6.     **end if**
7.     --- Attitude Control (LQR) ---
8.     $e_\eta \leftarrow [\phi_{des} - \phi, \theta_{des} - \theta, \psi_{des} - \psi]^T$;
9.     $U_2 \leftarrow k_{1,\phi} e_\phi - k_{2,\phi} p$;
10.     $U_3 \leftarrow k_{1,\theta} e_\theta - k_{2,\theta} q$;
11.     $U_4 \leftarrow k_{1,\psi} e_\psi - k_{2,\psi} r$;
12.     --- Altitude Control (PD) ---
13.     $e_z \leftarrow z_{des} - z$;
14.     $U_1 \leftarrow \frac{mg + K_{p,z} e_z - K_{d,z} \dot{z}}{\cos\phi \cos\theta}$;
15.     --- Control Allocation ---
16.     $U \leftarrow [U_1, U_2, U_3, U_4]^T$;
17.     $\Omega^2 \leftarrow M^{-1} U$;
18.     **for** (i←1) to 4 **do**
19.       $\omega_i \leftarrow \sqrt{\max(0, \Omega_i^2)}$;
20.     **end for**
21.     Pulish Ω to '/drone/motor_commands';
22.   **end procedure**
23. **end**

i. **Parameter Declaration and Retrieval**: Declare PID and LQR gain coefficients, desired altitude, attitude, and other configurable ROS 2 parameters, then fetch their initial values.
ii. **Communication Interface Setup**: Initialize multiple subscribers to receive state information from Simulink.
iii. **Control Loop Activation**: Create a high-frequency (50Hz) wall timer bound to the main control loop function controlLoop(). This timer-based approach replaces direct computation within callback functions, effectively decoupling sensor update rates from controller execution rates to ensure stable command output.

(2) State Estimation: Multiple subscriber callbacks asynchronously receive and update the vehicle's latest state in parallel. Specifically, imuCallback converts the incoming IMU attitude quaternion to Euler angles for controller utilization.

(3) Main Control Loop: The controlLoop() function, periodically triggered by the timer, serves as the node's core. Its execution flow is detailed in Algorithm 1.

Finally, the overall architecture of the proposed co-simulation framework is illustrated in Figure 3. This architecture adopts a fully decoupled design philosophy, where the physical model and control algorithm operate as independent entities in their respective optimal environments, communicating via the ROS 2 network.

- **Simulation Environment**: On an Ubuntu 24.04 host, a six-degree-of-freedom nonlinear dynamic model of the quadrotor is implemented using MATLAB/Simulink. This model serves as the plant, responsible for computing the aircraft's motion states in real-time based on received motor commands.
- **Controller**: Within an Ubuntu 24.04, an attitude controller is developed using C++ and the ROS 2 framework. Operating as a standalone ROS 2 node, this controller subscribes to the aircraft's state information, executes control algorithms, and publishes motor control commands.
- **Communication Middleware**: DDS serves as the communication bridge between the two platform. By defining standardized topics and message types, it enables stable, cross-platform and cross-language data exchange between the MATLAB/Simulink and ROS 2。

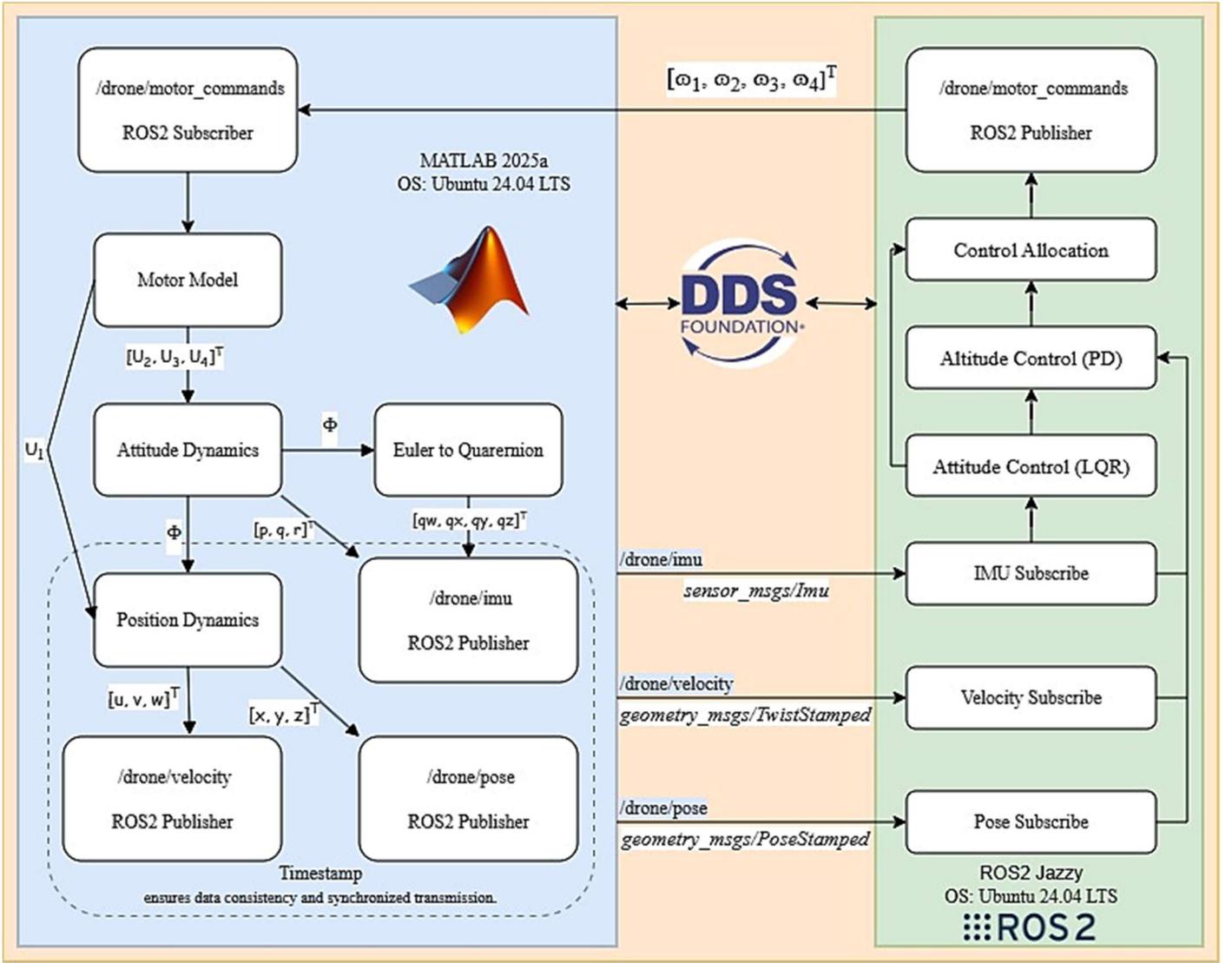

Figure 3 The co-simulation framework architecture

## IV. SIMULATION RESULTS AND DISCUSSION

To validate the effectiveness of the co-simulation framework proposed in Section III and evaluate the performance of the designed controller, this section presents a series of closed-loop simulation experiments. All experiments were conducted using the architecture shown in Figure 2, where the Simulink model operates with a fixed-step discrete solver at 100Hz, while the C++ controller node in ROS 2 executes periodically at 50Hz.

We design a series of step response test, to evaluate the tracking performance of the system in both position and attitude domains. Unless otherwise specified, the initial situation of quadrotor is set a static hover condition with all variables at zero. The total simulation time is configured as 15 seconds, to ensure the enough time for system to reach steady-state condition.

i. Parameters of the quadrotor

The parameters of quadrotor are configured as shown in following table 1:

Table 1 Parameters of the quadrotor

| Parameter | Symbol | Value | Unit |
|---|---|---|---|
| Mass | m | 1.96 | kg |
| Gravitational Acceleration | g | 9.81 | m/s² |
| Inertia X-axis | $I_{xx}$ | 0.0149 | kg·m² |
| Inertia Y-axis | $I_{yy}$ | 0.0153 | kg·m² |
| Inertia Z-axis | $I_{zz}$ | 0.0532 | kg·m² |
| Arm Length | L | 0.59 | m |
| Thrust Coefficient | $k_t$ | $2.06 \times 10^{-7}$ | N/(rad/s)² |
| Torque Coefficient | $k_m$ | $1.01 \times 10^{-10}$ | Nm/(rad/s)² |

ii. Altitude control test

To evaluate the performance of the altitude PD controller, we set the desired altitude to 10 meters via parameter callback on the control terminal, then recorded the quadrotor's altitude($z$) and vertical velocity($dz$) response curves, as shown in Figure 4.

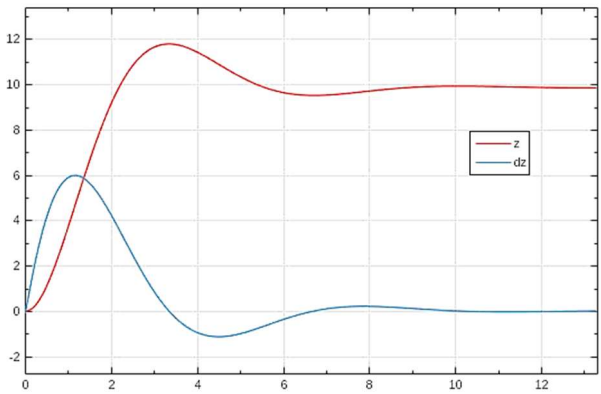

Figure 4 Step response of the altitude control system

Figure 4 illustrates the step response of the altitude control system. As shown in the plot, the quadrotor's altitude (red curve, z) responds promptly to the commanded reference input. The system reaches its peak at approximately 2.5 seconds with an overshoot of about 18% (11.8m). Subsequently, the PD controller effectively suppresses the oscillations, bringing the system to steady state within approximately 8 seconds. The altitude stabilizes at the desired 10-meter level with negligible steady-state error. The vertical velocity (blue curve, dz) exhibits the expected behavior: it increases rapidly during the initial phase and then decays to zero due to damping effects.

These results demonstrate that the designed PD controller achieves effective and stable altitude tracking performance.

iii. Attitude control tests

The performance of the LQR attitude tracking controller was independently evaluated for the roll, pitch, and yaw channels. Separate tests were conducted for each of the three attitude channels. During the roll test, the desired roll angle $\phi(phi)$ was set to 0.1 radians;

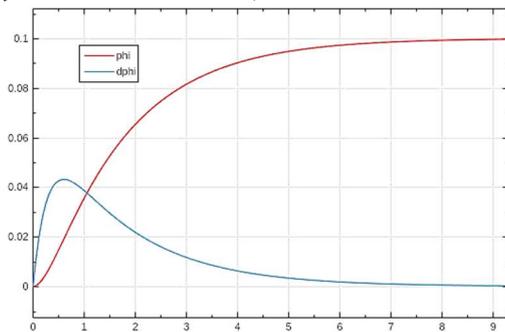

Figure 5 Step response of the rolling channel

Figure 5 illustrates the step response of the roll channel. As indicated by the red curve (phi), the system exhibits exceptionally rapid and smooth response characteristics with no observable overshoot. The quadrotor achieves approximately 95% of the desired attitude within 4 seconds and fully stabilizes at 0.1 radians after about 6 seconds, demonstrating zero steady-state error. The blue curve (dphi), representing roll angular velocity, displays an exemplary damped response, confirming that the LQR controller provides superior damping characteristics and effectively prevents oscillations.

for the pitch test, the desired pitch angle $\theta(theta)$ was set to 0.15 radians;

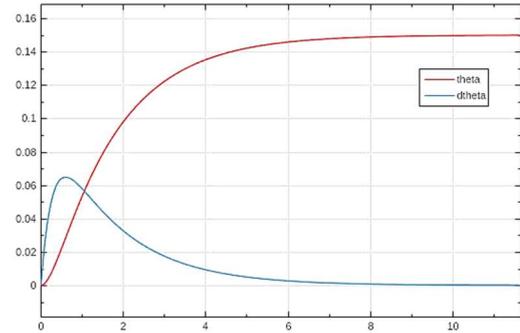

Figure 6 Step response of the pitch channel

Figure 6 illustrates the step response of the pitch channel, which exhibits remarkably similar characteristics to the roll channel, demonstrating rapid response and excellent performance without overshoot. Given that the moment of roll axis inertia $I_{xx}$ with the pitch axis inertia $I_{yy}$ closely approximates, the system achieves similar dynamic performance under analogous LQR weighting configurations, aligning perfectly with our theoretical predictions.

Finally in the yaw test, the desired yaw angle $\psi(psi)$ was set to 0.2 radians.

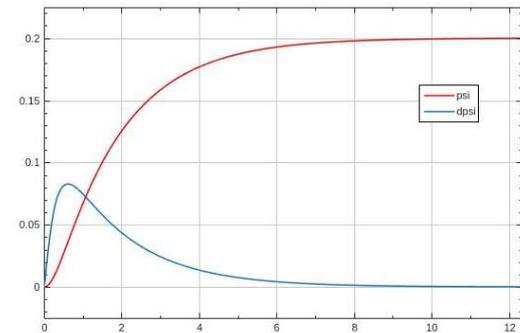

Figure 7 Step response of the yaw channel

Figure 7 illustrates the step response of the yaw channel. Compared to the roll and pitch channels, the yaw channel exhibits more slower response, taking approximately 8 seconds to approach steady-state. This behavior decided by the inherent physical characteristics of quadrotors, where equivalent control torque magnitudes produce smaller yaw angular accelerations, resulting in naturally slower responses. Despite the slower dynamics, the entire process remains remarkably stable, demonstrating neither overshoot nor oscillation, which confirms the L-QR controller's excellent adaptability to diverse channel dynamics. As above record the response curves of attitude angles and angular rates for each channel.

The simulation results clearly demonstrate the success of our designed hierarchical control strategy. The outer-loop PD altitude controller achieves stable altitude tracking, while the inner-loop LQR attitude controller delivers rapid, precise, and well-damped tracking performance across all three attitude channels. Zero steady-state error is achieved in all cases. These findings not only validate the effectiveness of our controller design but also confirm that the ROS 2 and Simulink co-simulation framework reliably supports complex closed-loop dynamic simulations, laying a solid foundation for the development and testing of more advanced control algorithms.

REFERENCES

[1] Arslan E, Suveren M, Moghaddam S T H. Ros gazebo and matlab/simulink co-simulation for cart-pole system: A framework for design optimization[C]//2023 7th International Symposium on Innovative Approaches in Smart Technologies (ISAS). IEEE, 2023: 1-7.

[2] Fernando H, De Silva A T A, De Zoysa M D C, et al. Modelling, simulation and implementation of a quadrotor UAV[C]//2013 IEEE 8th International conference on industrial and information systems. IEEE, 2013: 207-212.

[3] Majid A R A. Quadrotor UAV Kinematics and Dynamics: A Comprehensive Model with Suggested Control Techniques[C]//2024 International Conference on Modeling, Simulation & Intelligent Computing (MoSICom). IEEE, 2024: 335-340.

[4] Yao R. Design and Simulation of a Quadrotor UAV Control System Utilizing Sliding Mode Control Theory[C]//2024 13th International Conference of Information and Communication Technology (ICTech). IEEE, 2024: 258-262.

[5] Miura K, Tokunaga S, Horita Y, et al. Cosam: Co-simulation framework for ros-based self-driving systems and matlab/simulink[J]. Journal of Information Processing, 2021, 29: 227-235.

[6] Mizouri W, Najar S, Bouabdallah L, et al. Dynamic modeling of a quadrotor UAV prototype[M]//New Trends in Robot Control. Singapore: Springer Singapore, 2020: 281-299.